\documentclass[aps,pra,twocolumn]{revtex4}
\usepackage{graphicx}
\usepackage{bm}% bold math
\usepackage{amsmath}% equation formating
\usepackage{bbold}
\usepackage[utf8]{inputenc}
\usepackage{color}
\usepackage{soul}

\begin{document}
\title{Output squeezed radiation from dispersive ultrastrong light-matter coupling}
 
\author{ S. Fedortchenko$^1$, S. Huppert$^1$, A. Vasanelli$^1$, Y. Todorov$^1$, C. Sirtori$^1$, C. Ciuti$^1$, A. Keller$^2$,   T. Coudreau$^1$, and P. Milman$^1$ }

\affiliation{$^{1}$Laboratoire Mat\' eriaux et Ph\' enom\`enes Quantiques, Sorbonne Paris Cit\' e, Universit\' e Paris Diderot, CNRS UMR 7162, 75013, Paris, France}
\affiliation{$^{2}$Institut des Sciences Mol\'eculaires d'Orsay (ISMO), CNRS, Univ. Paris Sud, Universit\'e Paris-Saclay, F-91405 Orsay (France)}

\begin{abstract}
We investigate the output generation of squeezed radiation of a cavity photon mode coupled to another off-resonant bosonic excitation. By modulating in time their linear interaction, we predict high degree of output squeezing when the dispersive ultrastrong coupling regime is achieved, i.e., when the interaction rate becomes comparable to the frequency of the lowest energy mode. Our work paves the way to squeezed light generation in frequency domains where the ultrastrong coupling is obtained, e.g., solid-state resonators in the GHz, THz and mid-IR spectral range.
\end{abstract}
\vskip2pc

%\date{\today}
 
\vskip2pc 
\maketitle

\section{Introduction}

Experimentally demonstrated in 1965 \cite{Giordmaine}, Optical Parametric Oscillators (OPOs) have applications both for their classical \cite{Arslanov} and quantum properties. In quantum optics, they followed a path of fast advances since the eighties, with numerous theoretical contributions \cite{Milburn,Collett,Reynaud,Wolinsky,Fabre,Rubin}, as well as many experimental improvements \cite{Wu,Rarity,Wu2,Debuisschert,Mertz,Breitenbach}. These devices owe their growing interest to a notable feature: the generation of electromagnetic radiation with properties that cannot be predicted by classical Maxwell equations, namely squeezed states and entangled states of light, achieved also in the microwave domain \cite{Yurke,Beltran,Mallet,Eichler}. Apart from their fundamental interest, squeezed radiation enables metrology beyond the standard quantum limit \cite{Grote}, asymptotically reaching the ultimate Heisenberg limit \cite{Giovannetti} of noise reduction in a quantum measurement. Additionally, quantum information protocols with continuous variables rely on using squeezed states as a resource \cite{Braunstein,Cerf}, that can be entangled to form cluster states with more than hundreds of nodes \cite{Yokoyama,Chen,Wang}.

In this paper, we show how squeezed radiation can be created in linearly interacting light-matter systems by modulating in time the interaction between two off-resonant bosonic modes in an unusual dispersive ultrastrong coupling regime. Our results potentially enable the transposition of quantum optical effects to frequency domains unusual for the study of squeezed radiation, such as the THz and the mid-IR where the ultrastrong coupling regime can be experimentally observed.

The ultrastrong light-matter coupling, first suggested in Ref.\cite{Ciuti2}, became achievable in recent years. Its simplest definition is when the Rabi frequency, the interaction strength between the electromagnetic field and a material system, reaches the order of magnitude of the resonance frequencies of the system, thus breaking the extensively studied rotating-wave approximation. Among examples of experimental systems where this peculiar regime can be observed are cavity embedded doped semiconductors~\cite{Gunter,Todorov2,Askenazi} and superconducting devices~\cite{Niemczyk}. The failure of the rotating-wave approximation to describe the coupling between two quantum systems is usually studied in the resonant regime where both systems have the same characteristic frequency. Here we consider an ultrastrongly coupled device in the unusual off-resonant regime. Resonance is achieved by modulating in time the coupling between both bosonic modes, which leads to the emergence of the ultrastrong coupling regime. We show that this situation bears similarities with the case of an OPO, allowing to strongly squeeze radiation, yet with significant differences. An interesting aspect brought to light here, which is a direct consequence of the ultrastrong coupling, is the existence of nontrivial resonance conditions for which the noise spectrum reveals squeezing while displaying peculiar symmetry properties.

Before considering the model used in our work, we briefly recall some essential principles of an OPO, since its non linearities will help us providing some intuition on the obtained results. In a standard optical parametric oscillator a nonlinear crystal is placed inside a cavity and driven by an intense pump field. A nonlinear interaction is thus mediated by the crystal, involving the pump and two other modes of the electromagnetic field, the so-called signal and idler fields. As a result of the parametric down conversion process, one obtains a two mode (signal and idler) squeezed state. The pump field is usually treated classically, and the interaction Hamiltonian of the process, written in a frame rotating at the pump frequency, is
\begin{equation}
\hat{H}_{OPO} \propto i \, \frac{g \, \alpha_{\text{Pump}} }{2} \,  (  \hat{a}_{s}^{\dagger} \hat{a}_{i}^{\dagger} - \hat{a}_{s} \hat{a}_{i}),
\label{HOPO}
\end{equation}
where $\hat{a}_{s}(\hat{a}_{s}^{\dagger})$ is the annihilation(creation) operator of the signal, and $\hat{a}_{i}(\hat{a}_{i}^{\dagger})$ is the annihilation(creation) operator of the idler. $\alpha_{\text{Pump}}$ is the amplitude of the pump field, assumed real here, and $g$ is the coupling of the nonlinear process. This interaction can lead to the production of two--mode squeezed radiation in the non--degenerate case ($\omega_{s} \neq \omega_{i}$) and single mode squeezing in the degenerate one ($\omega_{s} = \omega_{i}$).

\begin{figure}[b!]
\centering
\includegraphics[width=0.47\textwidth]{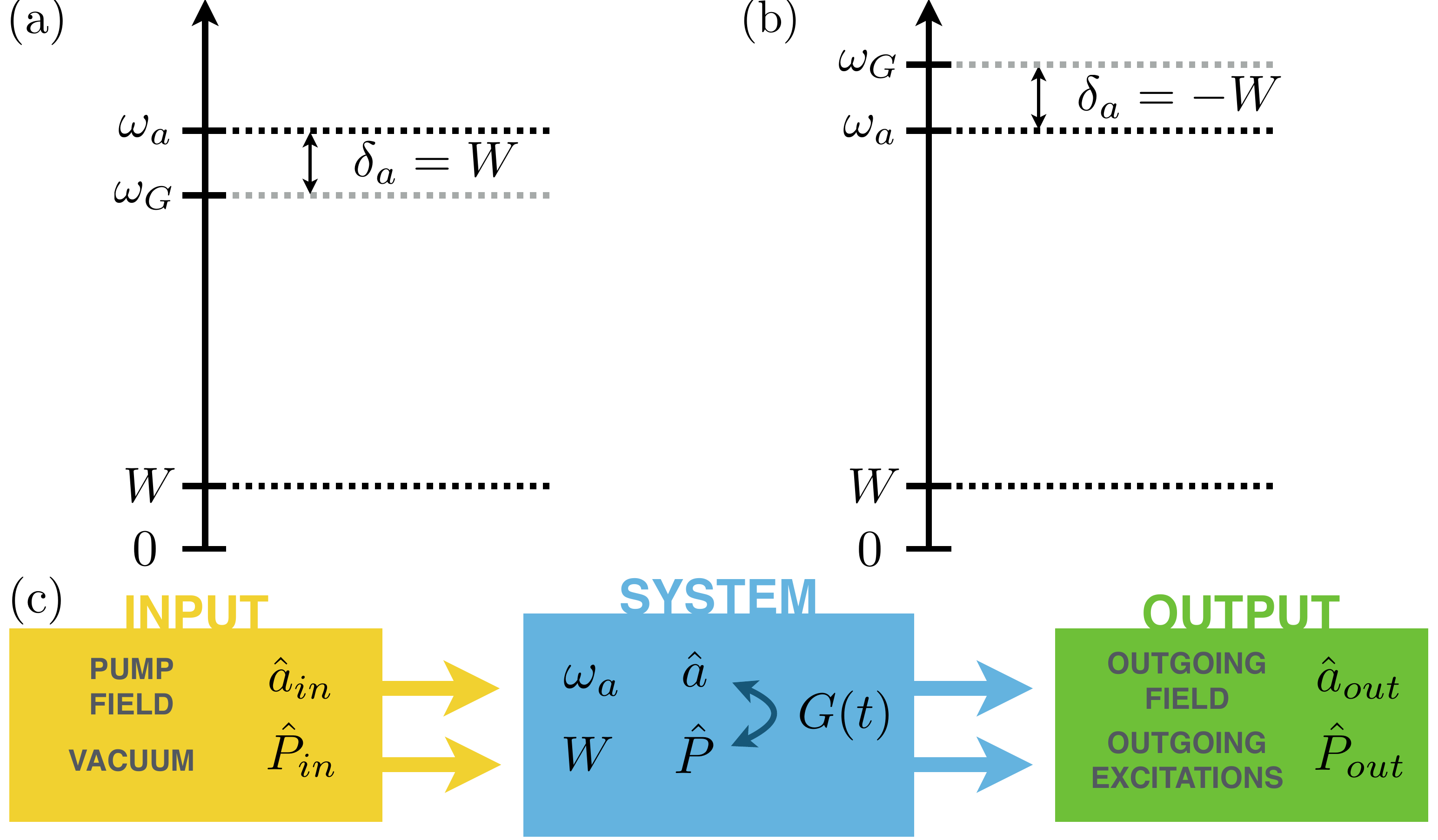}
\caption{(a) Resonance condition of the system frequencies where squeezing cannot be observed in the strong coupling regime : $\delta_{a} = W$, where $\delta_a=\omega_a-\omega_G$. (b) Another resonance condition studied here, corresponding to the intuitive way to generate two--mode squeezing in the strong coupling regime : $\delta_{a} = - W$. (c) Input-output scheme where the input, a coherent pump field, is sent on the system, and the output, a field coming out of the system, is analysed.}
\label{FigureSystem}
\end{figure}

\section{Modelling the dispersive ultrastrong coupling}

Let us consider the Hamiltonian (in $\hbar \equiv 1$ units) of two coupled bosonic fields, one describing a cavity mode, while the other describing a bosonic matter excitation or another cavity mode,
\begin{equation}
\hat{H}_{S} = \omega_{a} \, \hat{a}^{\dagger} \hat{a} + W \, \hat{P}^{\dagger} \hat{P} +  i \, G(t) (  \hat{a}^{\dagger} - \hat{a} ) ( \hat{P}^{\dagger} + \hat{P}),
\label{HS}
\end{equation}
where $\hat{a}(\hat{a}^{\dagger})$ is the annihilation(creation) operator of a bosonic mode at frequency $\omega_{a}$, and where $\hat{P}(\hat{P}^{\dagger})$ is the annihilation(creation) operator of a non-degenerate bosonic excitation with frequency $W$. $G (t) = G_{0} + G_{\text{mod}} \cos{(\omega_{G} t)}$ represents the coupling between the two bosonic fields, modulated at a frequency $\omega_{G}$.  In the absence of modulation, the Rabi frequency is simply $G_{0}$. This non modulating regime can only lead to squeezing if one of the fields obeys non--linear dynamics such as in Ref.~\cite{Stassi}.  As mentioned before, the two bosonic modes are non degenerate, and fulfill a resonance condition with respect to $\omega_G$, as shown in Fig.~\ref{FigureSystem}(a) and \ref{FigureSystem}(b). A sinusoidal modulation of the Rabi coupling was also considered in Ref.~\cite{DeLiberato} for the study of quantum vacuum radiation, and in Ref.~\cite{Felicetti}, as a mean to entangle artificial atoms.  Also, this Hamiltonian was considered in the strong coupling regime as a strategy to create two--mode squeezed states in the microwave range in the non--degenerate case \cite{Wilson}, and one mode squeezing in the degenerate one \cite{Beltran}. In such a coupling regime, the analogy between the Hamiltonian (\ref{HS}) and the usual OPO model is direct, if one considers that the coupling modulation in Eq.~(\ref{HS}) plays the role of the pump field in Eq.~(\ref{HOPO}). We stress that this well studied and understood regime is not the scope of the present contribution. 

By using the unitary transformation $\hat{U} (t) = e^{i \omega_{G} \hat{a}^\dagger \hat{a} t}$, we now move to a frame rotating at $\omega_{G}$, with time dependent terms oscillating at $\omega_{G}$ or at $2 \, \omega_{G}$. By working in a regime where $G_{0}, G_{\text{mod}} \ll \omega_{G}$ and $W < \omega_{G}$, these terms become fast oscillating ones, and can therefore be eliminated by a well justified rotating wave approximation (see Appendix). In the rotating frame, the remaining term of the light-matter interaction allow us to write the effective Hamiltonian
\begin{equation}
\hat{H}_{\text{eff}} = \delta_{a} \, \hat{a}^{\dagger} \hat{a} + W \, \hat{P}^{\dagger} \hat{P} +  i \, G_{\text{mod}} (  \hat{a}^{\dagger} - \hat{a} ) ( \hat{P}^{\dagger} + \hat{P})/2,
\label{Heff}
\end{equation}
where $\delta_{a} = \omega_{a} - \omega_{G}$.

In the frame rotating at frequency $\omega_{G}$, the standard procedure of Input-Output theory  can be used \cite{Ciuti,Gardiner,GardinerBook,WallsBook}, as illustrated in Fig.~\ref{FigureSystem}(c). Since in the considered dispersive ultrastrong coupling regime the modulation frequency in much larger than the modulation amplitude, $G_{\text{mod}} \ll \omega_{G}$, the relevant spectral band in the dynamics remains around the central frequency $\omega_{G}$ and involve only in a negligible way the zero and negative frequencies. In such a configuration, we have explicitly verified that the ultrastrong coupling Input-Output theory which mandatorily requires frequency-dependent damping rates \cite{Ciuti}, simplifies and reduces to the standard one. It leads to  the following quantum stochastic equations of motion in frequency space:
\begin{align}
-i \, \omega \, \hat{a} (\omega)  = & - i \, \big( \delta_{a} + \Gamma_{a} (\omega) \big) \hat{a} (\omega) + G_{\text{mod}} \big( \hat{P}^{\dagger} (-\omega) + \nonumber \\
& + \hat{P} (\omega) \big) -\sqrt{2 \pi} \kappa_{a} (\omega) \hat{a}_{\text{IN}} (\omega) \label{Heisenberg1} \\
-i \, \omega \, \hat{P} (\omega)  = & - i \, \big( W + \Gamma_{P} (\omega) \big) \hat{P} (\omega) + G_{\text{mod}} \big( \hat{a}^{\dagger} (-\omega) - \nonumber \\
& - \hat{a} (\omega) \big) - \sqrt{2 \pi} \kappa_{P} (\omega) \hat{P}_{\text{IN}} (\omega),
\label{Heisenberg2}
\end{align}
$\Gamma_{i} (\omega)$ and the terms proportional to $\kappa_{i} (\omega)$ refer respectively to the dissipation kernels and to the Langevin forces, originated from the coupling of each subsystem to its own environment. In the Input-Output formalism, the Langevin forces contain the inputs, $\hat{a}_{\text{IN}} (\omega)$ and $\hat{P}_{\text{IN}} (\omega)$, considered for each subsystem as a mode of its respective environment. The average part of $\hat{a}_{\text{IN}} (\omega)$ contains the input coherent drive, whereas its fluctuations contain the noise entering the cavity. For simplicity, we assume that the system is that zero temperature, so that the fluctuations of $\hat{P}_{\text{IN}} (\omega)$ correspond to vacuum noise. It is important to stress again the originality of the regime studied here with respect to the ones previously studied theoretically and experimentally: in the present work, the two bosonic fields are not in resonance, as usually considered. Thus, the coupling is considered as ultrastrong if $G_{\text{mod}}$ is a significant fraction of $W$, which has been arbitrarily chosen to be the lower frequency mode. Therefore, since the two modes are strongly detuned, we call this regime as dispersive ultrastrong coupling. Note that in the absence of modulation, only intracavity squeezed radiation can be produced \cite{Ciuti}, since noise reduced radiation emission does not conserve energy. As will be seen in the following, the regime considered here will lead to an exotic output noise spectrum displaying squeezing at unusual resonance frequencies. 

\begin{figure}[b!]
\centering
\includegraphics[width=0.48\textwidth]{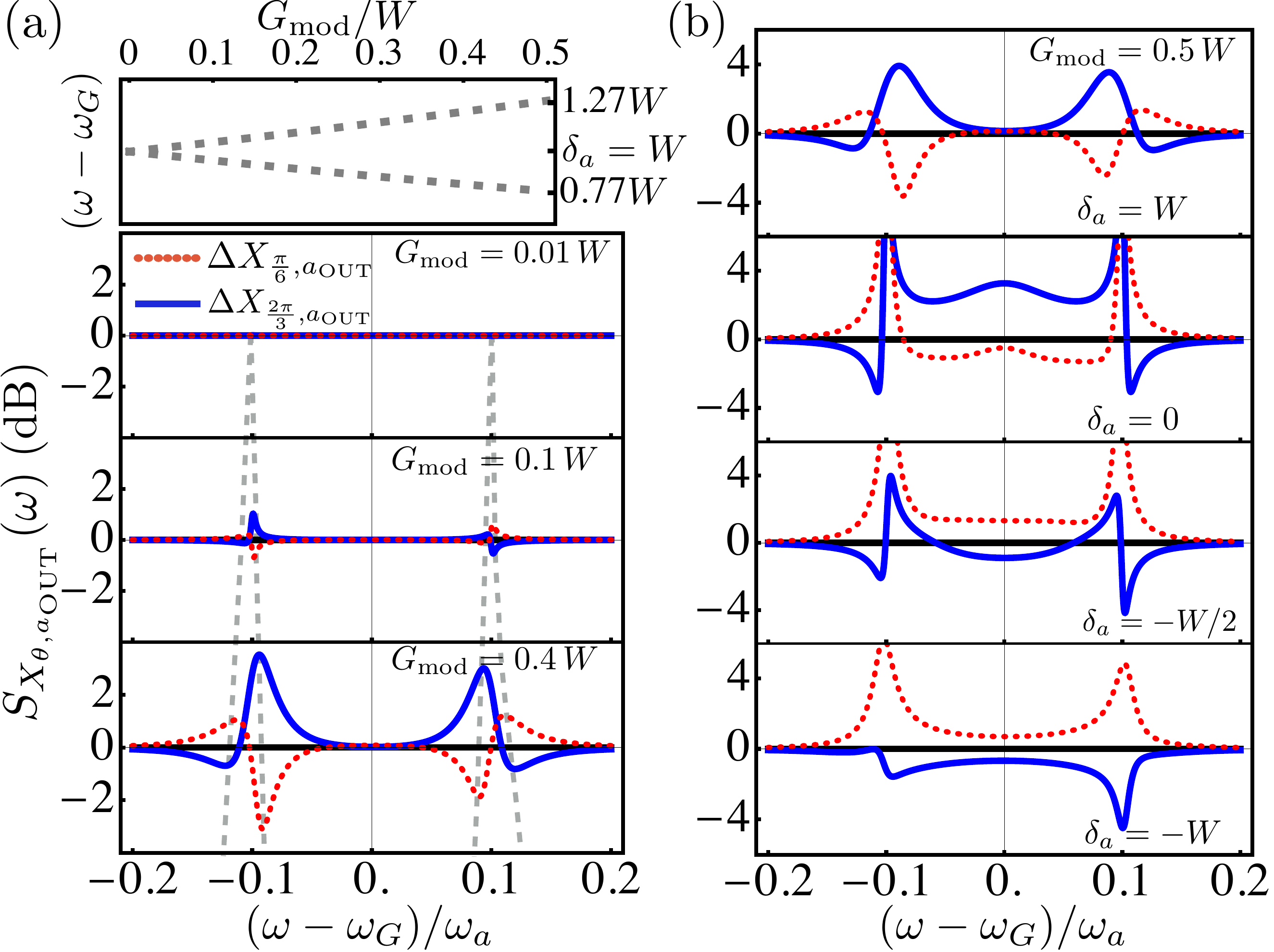}
\caption{ Energy splitting from the Hamiltonian (\ref{Heff}), as a consequence of the ultrastrong coupling regime and noise spectra $S_{X_{\pi/6,a_{\text{OUT}}}} (\omega)$ (red dotted) and $S_{X_{2\pi/3,a_{\text{OUT}}}} (\omega)$ (blue full) versus the analysis frequency $\omega' = \omega - \omega_{G}$ in the rotating frame. The parameters are : $\Gamma_{i} (\omega) = \gamma_{i}/2$ ; $\kappa_{i} (\omega) = \sqrt{ \gamma_{i} / 2 \pi}$ ; $W= 0.1 \, \omega_{a}$; $\gamma_{a} = \omega_{a}/10$; $\gamma_{P} = W/30$. (a) Upper panel : polaritonic splitting. Lower panels : $\omega_{G} = 0.9 \, \omega_{a}$ and from top to bottom : $G_{\text{mod}}=0.01 \, W$; $G_{\text{mod}}=0.1 \, W$; $G_{\text{mod}}=0.4 \, W$. (b) $G_{\text{mod}}=0.5 \, W$ and from top to bottom : $\delta_{a}= W$; $\delta_{a}= 0$; $\delta_{a}= -W/2$; $\delta_{a}= - W$, with $\delta_{a} = \omega_{a}-\omega_{G}$.  }
\label{FigurenaOUTSaOUT}
\end{figure}

\section{Results}

One can solve the Heisenberg equations (\ref{Heisenberg1}) and (\ref{Heisenberg2}) and obtain the relation between the output field $\hat{a}_{\text{OUT}} (\omega)$ and the input field $\hat{a}_{\text{IN}} (\omega)$. With this input-output relation one can compute the noise spectrum (in dB) of the quadrature $\hat{X}_{\theta} = (e^{-i \theta} \hat{a} + e^{i \theta} \hat{a}^\dagger)/\sqrt{2}$, defined by $S_{X_{\theta}} = 10 \log_{10}{( \langle \Delta \hat{X}_{\theta}^2 \rangle / \langle \Delta \hat{X}_{coh}^2 \rangle)}$. We use the definition $\langle \Delta \hat{X}_{\theta}^2 \rangle = \langle \hat{X}_{\theta}^2 \rangle - \langle \hat{X}_{\theta} \rangle^2$, and $\langle \Delta \hat{X}_{coh}^2 \rangle$ corresponds to the noise of a coherent state. With this definition, the standard quantum limit corresponds to 0 dB, and contributions below this limit correspond to squeezing. We show in Fig. \ref{FigurenaOUTSaOUT} for different values of $G_{\text{mod}}$ and of $\omega_{G}$ the noise spectra $S_{X_{\theta}}$ as a function of the analysis frequency (or pump frequency) $\omega' = \omega - \omega_{G}$ ($i.e.$, in the rotating frame), for the orthogonal quadratures $\hat{X}_{\theta=\pi/6}$  and $\hat{X}_{\theta=2\pi/3}$.  

The upper panel in Fig. \ref{FigurenaOUTSaOUT}(a) displays the polaritonic splitting, \textit{i.e.}, the eigenenergies of the effective Hamiltonian (\ref{Heff}) as a function of $G_{\text{mod}}$, occuring in our model when ultrastrong light-matter interaction takes place. The other panels serve to illustrate this behavior, as we increase the modulation amplitude $G_{\text{mod}}$ from top to bottom. We start by discussing the basic physical features involved in the noise spectra of  quadrature $\Delta \hat{X}_{\pi/6,a_{\text{OUT}}}$. The behavior of $\Delta \hat{X}_{2\pi/3,a_{\text{OUT}}}$ will follow the same principles and will be rapidly discussed afterwards. In the lower panels of Fig. \ref{FigurenaOUTSaOUT}(a) one notices that the noise spectra can display squeezing for certain frequencies, corresponding to the system resonances. By analyzing the dependency of squeezing with $G_{\text{mod}}$, we notice that noise reduction below the standard quantum limit is a direct consequence of the ultrastrong coupling regime, and is shown to be more significant as we further increase the modulation amplitude $G_{\text{mod}}$, reaching -3 dB for $G_{\text{mod}}= 0.4 \, W$.  In particular, we can notice that in the lower panels of Fig. \ref{FigurenaOUTSaOUT}(a), the region where $\omega > \omega_G$ leads to the resonance of the  $\hat a \hat P^{\dagger}$ ($\hat a^{\dagger} \hat P$) terms in Eq. (\ref{Heff}), because in the interaction picture these terms are oscillating at $\delta_{a} - W$. In the strong coupling regime, there is no squeezing, since the  $\hat a^{\dagger}\hat P^{\dagger}$($\hat a \hat P$) terms are rapidly oscillating at $\delta_{a} + W$ and $G_{\text{mod}} \ll W$. However, as shown in the lower panels, squeezing appears with the emergence of the ultrastrong limit, for which these terms cannot be neglected and we observe the Rabi splitting of polaritons. As a guide to the eye, we reproduced the polaritonic splitting of the upper panel of Fig. \ref{FigurenaOUTSaOUT}(a) in dotted gray lines in the lower panels. Interestingly, the frequency where squeezing is observed for $\omega > \omega_G$ follows well the new polaritonic eigenfrequencies. As for the $\omega < \omega_G$ region of the lower panels of Fig. \ref{FigurenaOUTSaOUT}(a), a different type of resonance condition is observed. This new resonance condition is at the origin of the asymmetry of the noise spectra with respect to $\omega = \omega_G$. For $\omega < \omega_G$, the terms $\hat a^{\dagger} \hat P^{\dagger}$ ($\hat a \hat P$) of  Eq. (\ref{Heff}) are closer to resonance, which occurs for $\delta_{a} + W = 0$, and thus the spectra perfectly follows the shape of the polaritonic splitting. Notice that these terms correspond to the standard OPO regime, described by Eq.~(\ref{HOPO}). Nevertheless, it is important to stress that the observed noise reduction is a single mode property rather than a two mode one, as is the case of the OPO Hamiltonian in the non--degenerate regime. 

The above discussion also explains the qualitative behavior of the noise spectra of the quadrature  $\Delta \hat{X}_{2\pi/3,a_{\text{OUT}}}$. Therefore, we notice that while one polariton mode displays squeezing in one quadrature, the other displays it in the orthogonal one. 

We stress that values above $0.5 \, W$ were not considered in the present model of $G_{\text{mod}}$, since we kept the constraint $G_{0} + G_{\text{mod}} \leq W$ and $G_{0} \geq G_{\text{mod}}$. Nevertheless, according to the considered experimental systems, higher values of the modulation amplitude can be considered, potentially leading to higher squeezing. The particular features of possible experimental implementations will be briefly discussed in this manuscript and detailed elsewhere. 

In Fig. \ref{FigurenaOUTSaOUT}(b) we show the dependency of the squeezing spectrum  with $\omega_{G}$. The upper panel correspond to $\delta_{a}=W$. The second panel from the top stands for $\delta_{a}=0$, where both spectra are symmetric with respect to the rotating frame origin. The third panel from the top is for $\delta_{a}=-W/2$ and the bottom one is for $\delta_{a}=-W$, thus it is on the positive side of the panel that we see the consequence of the  terms $\hat{a}^\dagger \hat{P}^\dagger$ and $\hat{a} \, \hat{P}$, in opposition with the first panel.  The polaritonic splitting becomes difficult to see since in this regime only the upper polariton leads to squeezing, and the losses are significant (they correspond to an experimental system introduced hereafter). By increasing $\omega_{G}$ and $\vert \delta_{a} \vert$ until $\delta_{a}=-W$, one can observe more than -4 dB of squeezing, as shown in the bottom panel. 

\section{Discussion on the experimental implementation of the model}

We now discuss some experimental systems to which our results can be directly applied or safely adapted. The basic requirements such systems must satisfy is the realization of the ultrastrong coupling regime, of its time modulation and, finally, the detection of amplitude and phase information of non--classical radiation. Two intrinsically different physical systems that are potentially good candidates for satisfying these conditions are superconducting circuits and  cavity embedded semiconductor quantum wells. We will briefly discuss both of them, the achievability of the required conditions and the technological progress enabled by our results. 

The ultrastrong limit has been achieved experimentally in cavity embedded semiconductor quantum wells \cite{Todorov,Delteil}, where one bosonic mode corresponds to the electromagnetic field in the cavity while the other is a collective excitation of a confined electron gas, called the multisubband plasmon. For the derivation of the Hamiltonian (\ref{HS}) for such systems in the case $G(t)=G_{0}$, we refer to Refs.~\cite{CohenBook,Todorov3,Pegolotti}. Note that this Hamiltonian is written in the Power-Zienau-Woolley representation of quantum electrodynamics \cite{CohenBook}, which allows one to take into account for the Coulomb interactions and for the quadrupolar term in a very convenient way. Matching the experimental achievability for such systems \cite{Delteil,Todorov}, the parameters on Fig. \ref{FigurenaOUTSaOUT} would correspond to a plasmon of few THz while the system is pumped at few tens of THz. Correspondingly, the coupling $G_{\text{mod}}$ should be of the order of a few THz, allowing to reach the ultrastrong limit considered here. Alternatively, one could also observe a situation where the plasmon is at few tens of THz and therefore the system is pumped at mid-Infrared frequencies. On the experimental implementation of the time modulation in the Rabi coupling, crucial for the generation of squeezing, first demonstrations of time dependent light matter interaction turned on in an ultrafast timescale were observed in Refs.\cite{Gunter,Porer}. Even if this is not the exact time dependency discussed in this manuscript, these realizations indicates the possibility of implementing different time dependent light-matter couplings, which is the essential ingredient to the emergence of the ultrastrong limit as described in our model. In addition, the theoretical study of the generation of squeezed states within the framework of already experimentally observed time dependent couplings  is an interesting perspective for a future work, but not in the scope of this paper. Regarding squeezing detection, although the state of the art of homodyne detection in the THz and mid-IR ranges is not as advanced as the one involving optical frequencies, experimental demonstrations with the retrieving of both amplitude and phase information exist \cite{Riek}. One can thus be optimistic about the application of that experimental result to the detection of squeezing. We stress that although it is possible to produce THz electromagnetic waves for such frequencies by using regular non--degenerate OPOs, the generation of single mode squeezed states in this frequency range remains unexploited and unachievable in such experiments \cite{Kawase,Edwards,Molter}. So far, to the best of our knowledge, apart from the present proposal, there are no experimental realizations nor other theoretical proposals for non--classical radiation generation in the THz and the mid-IR domains, even though there is an emerging interest of the community in the study of quantum optics in such frequency ranges \cite{Benea-Chelmus}. 

The oscillation in time of the Rabi coupling used in this work has been experimentally achieved in superconducting systems \cite{Wilson,Bialczak}, where the ultrastrong coupling limit has also been reached \cite{Niemczyk}. In such systems, the two bosonic fields studied here would represent the quantized field of two non--degenerate coplanar cavities coupled by a non--linear superconducting element, as a superconducting quantum interference device, an architecture demonstrated in Ref.~\cite{Eichler}. In this type of systems, single mode squeezed radiation can be created for frequencies from the few to tens of GHz, and there are well developed experimental techniques for both homodyne \cite{daSilva,Petersson} and heterodyne \cite{Yin,Ibarcq} detections.

\section{Conclusion} 

In conclusion, we have shown how the dispersive ultrastrong interaction, an original coupling regime of bosonic systems, can lead to single mode squeezed radiation emission, when modulated in time, in a wide variety of frequency ranges, so far unexplored for quantum optical purposes. We provide an extensive discussion and interpretation of our results that can be applied to experimental systems at very different energies.  We specifically discussed two different experimental set-ups as suitable candidates to demonstrate, with current technology, our results. With the perspective of improvement of the detection and of the capability in modulating the light-matter coupling on a very short timescale, squeezed light in these frequencies may be used for spectroscopy \cite{Yasui}, interferometry \cite{Rakic}, or precision measurements \cite{Hoshina}. Finally, a natural perspective of our results is investigating the possibility of  two mode squeezed states generation in the ultrastrong coupling regime.

\section*{Acknowledgments}

This work was supported by the French Agence Nationale de la Recherche (ANR COMB project, Grant No. ANR-13-BS04-0014 and Grant No. ANR-14-CE26-0023-01) and by Labex SEAM. The authors acknowledge C. Fabre, J. Restrepo, and F. Portier for helpful discussions.

\section*{Appendix: Derivation of the effective system Hamiltonian}

We start from the system Hamiltonian~(\ref{HS}) 
\begin{equation}
\hat{H}_{S} = \omega_{a} \, \hat{a}^{\dagger} \hat{a} + W \, \hat{P}^{\dagger} \hat{P} +  i \, G(t) (  \hat{a}^{\dagger} - \hat{a} ) ( \hat{P}^{\dagger} + \hat{P}).
\label{HSAppendix}
\end{equation}
By applying the unitary transformation $\hat{U} (t) = e^{i \omega_{G} \hat{a}^\dagger \hat{a} t}$ to Eq.~(\ref{HSAppendix}), we obtain
\begin{align}
\hat{H}_{S}' & = \delta_{a} \, \hat{a}^{\dagger} \hat{a} + W \, \hat{P}^{\dagger} \hat{P} + \nonumber \\
&  + i \, G (t) (  \hat{a}^{\dagger} e^{i \omega_{G} t} - \hat{a} e^{-i \omega_{G}  t} ) ( \hat{P}^{\dagger} + \hat{P}),
\label{HS2Appendix}
\end{align}
where $\delta_{a}= \omega_{a} -\omega_{G}$. By recalling that $G (t) = G_{0} + G_{\text{mod}} \cos{(\omega_{G} t)}$, we obtain
\begin{align}
\hat{H}_{S}' & = \delta_{a} \, \hat{a}^{\dagger} \hat{a} + W \, \hat{P}^{\dagger} \hat{P} + \nonumber \\
&  + i \, G_{0} (  \hat{a}^{\dagger} e^{i \omega_{G} t} - \hat{a} e^{-i \omega_{G}  t} ) ( \hat{P}^{\dagger} + \hat{P})+ \nonumber \\
&  + i \, G_{\text{mod}} (  \hat{a}^{\dagger} e^{2 i \omega_{G} t} - \hat{a} e^{-2 i \omega_{G}  t} ) ( \hat{P}^{\dagger} + \hat{P})/2+ \nonumber \\
&  + i \, G_{\text{mod}}  (  \hat{a}^{\dagger} - \hat{a}) ( \hat{P}^{\dagger} + \hat{P})/2.
\label{HS3Appendix}
\end{align}
We work in a regime where $W \sim \vert \delta_{a} \vert$, $W < \omega_{G}$ and  $G_{0}, G_{\text{mod}} \ll \omega_{G}$, meaning that the terms oscillating at $\omega_{G}$ and $2 \omega_{G}$ can be safely neglected using the rotating wave approximation. Consequently, we end up with the effective Hamiltonian seen in Eq.~(\ref{Heff}),
\begin{equation}
\hat{H}_{\text{eff}} = \delta_{a} \, \hat{a}^{\dagger} \hat{a} + W \, \hat{P}^{\dagger} \hat{P} +  i \, G_{\text{mod}} (  \hat{a}^{\dagger} - \hat{a} ) ( \hat{P}^{\dagger} + \hat{P})/2.
\label{HeffAppendix}
\end{equation}


\begin{thebibliography}{99}
\bibitem{Giordmaine} J. A. Giordmaine and R. C. Miller, Phys. Rev. Lett. {\bf 14}, 973 (1965).

\bibitem{Arslanov} D. D. Arslanov, M. P. P. Castro, N. A. Creemers, A. H. Neerincx, M. Spunei ; J. Mandon, S. M. Cristescu, P. Merkus, F. J. M. Harren, J. Biomed. Opt. {\bf 18}, 107002 (2013).

\bibitem{Milburn}  G. J. Milburn and D. F. Walls, Opt. Commun. {\bf 39}, 401 (1981).

\bibitem{Collett} M. J. Collett and D. F. Walls, Phys. Rev. A {\bf 32}, 2887 (1985).

\bibitem{Reynaud} S. Reynaud, C. Fabre, and E. Giacobino, J. Opt. Soc. Am. B {\bf 4}, 1520 (1987).

\bibitem{Wolinsky} M. Wolinsky and H. J. Carmichael, Phys. Rev. Lett. {\bf 60}, 1836 (1988).

\bibitem{Fabre} C. Fabre, E. Giacobino, A. Heidmann, L. Lugiato, S. Reynaud, M. Vadacchino and W. Kaige, Quantum Opt. {\bf 2}, 159 (1990).

\bibitem{Rubin} M. H. Rubin, D. N. Klyshko, Y. H. Shih, and A. V. Sergienko, Phys. Rev. A {\bf 50}, 5122 (1994).

\bibitem{Wu} L.-A. Wu, H. J. Kimble, J. L. Hall and Huifa Wu, Phys. Rev. Lett. {\bf 57}, 2520 (1986).

\bibitem{Rarity} J. G. Rarity, P. R. Tapster and E. Jakeman, Opt. Commun. {\bf 62}, 201 (1987).

\bibitem{Wu2} L.-A. Wu, M. Xiao, and H. J. Kimble, J. Opt. Soc. Am. B {\bf 4}, 1465 (1987).

\bibitem{Debuisschert} T. Debuisschert, S. Reynaud, A. Heidmann, E. Giacobino and C. Fabre, Quantum Opt. {\bf 1}, 3 (1989).

\bibitem{Mertz} J. Mertz, A. Heidmann, C. Fabre, E. Giacobino and S. Reynaud, Phys. Rev. Lett. {\bf 64}, 2897 (1990).

\bibitem{Breitenbach} G. Breitenbach, T. M\"uller, S.F. Pereira, J.-P. Poizat, S. Schiller and J. Mlynek, J. Opt. Soc. Am. B {\bf 12}, 2304 (1995).

\bibitem{Yurke} B. Yurke, P. G. Kaminsky, R. E. Miller, E. A. Whittaker, A. D. Smith, A. H. Silver, and R. W. Simon, Phys. Rev. Lett. {\bf 60}, 764 (1988).

\bibitem{Beltran} M. A. Castellanos-Beltran, K. D. Irwin, G. C. Hilton, L. R. Vale and K. W. Lehnert, Nature Phys. {\bf 4}, 929 (2008).

\bibitem{Mallet} F. Mallet, M. A. Castellanos-Beltran, H. S. Ku, S. Glancy, E. Knill, K. D. Irwin, G. C. Hilton, L. R. Vale, and K. W. Lehnert, Phys. Rev. Lett. {\bf 106}, 220502  (2011).

\bibitem{Eichler} C. Eichler, Y. Salathe, J. Mlynek, S. Schmidt, and A. Wallraff, Phys. Rev. Lett. {\bf 113}, 110502 (2014).

\bibitem{Grote} H. Grote, K. Danzmann, K. L. Dooley, R. Schnabel, J. Slutsky and H. Vahlbruch, Phys. Rev. Lett. {\bf 110}, 181101 (2013).

\bibitem{Giovannetti} V. Giovannetti, S. Lloyd, L. Maccone, Science {\bf 306}, 1330 (2004).

\bibitem{Braunstein} S. L. Braunstein and H. J. Kimble, Phys. Rev. Lett. {\bf 80}, 869 (1998).

\bibitem{Cerf} N. J. Cerf, M. L\'evy and G. Van Assche, Phys. Rev. A {\bf 63}, 052311 (2001).

\bibitem{Yokoyama} S. Yokoyama, R. Ukai, S. C. Armstrong, C. Sornphiphatphong, T. Kaji, S. Suzuki, J.-i. Yoshikawa, H. Yonezawa, N. C. Menicucci and A. Furusawa, Nat. Phot. {\bf 7}, 982 (2013).

\bibitem{Chen} M. Chen, N. C. Menicucci, and O. Pfister, Phys. Rev. Lett. {\bf 112}, 120505 (2014).

\bibitem{Wang} P. Wang, W. Fan, and O. Pfister, arXiv:1403.6631v2 (2014).

\bibitem{Ciuti2} C. Ciuti, G. Bastard, I. Carusotto, Phys. Rev. B {\bf 72}, 115303 (2005).

\bibitem{Gunter} G. G\"unter, A. A. Anappara, J. Hees, A. Sell, G. Biasiol, L. Sorba, S. De Liberato, C. Ciuti, A. Tredicucci, A. Leitenstorfer and R. Huber, Nature {\bf 458}, 178 (2009).

\bibitem{Todorov2} Y. Todorov, A. M. Andrews, R. Colombelli, S. De Liberato, C. Ciuti, P. Klang, G. Strasser, and C. Sirtori, Phys. Rev. Lett. {\bf 105}, 196402 (2010).

\bibitem{Askenazi} B. Askenazi, A. Vasanelli, A. Delteil, Y. Todorov, L. C. Andreani, G. Beaudoin, I. Sagnes, and C. Sirtori, New J. Phys. {\bf 16}, 043029 (2014).

\bibitem{Niemczyk} T. Niemczyk, F. Deppe, H. Huebl, E. P. Menzel, F. Hocke, M. J. Schwarz, J. J. Garcia-Ripoll, D. Zueco, T. H\"ummer, E. Solano, A. Marx and R. Gross, Nature Phys. {\bf 6}, 772 (2010).

\bibitem{Stassi} R. Stassi, S. Savasta, L. Garziano, B. Spagnolo, F. Nori, arXiv:1509.09064 (2015).

\bibitem{DeLiberato} S. De Liberato, C. Ciuti, and I. Carusotto, Phys. Rev. Lett. {\bf 98}, 103602 (2007).

\bibitem{Felicetti} S. Felicetti, M. Sanz, L. Lamata, G. Romero, G. Johansson, P. Delsing, and E. Solano, Phys. Rev. Lett. {\bf 113}, 093602 (2014).

\bibitem{Wilson} C. M. Wilson, G. Johansson, A. Pourkabirian, M. Simoen, J. R. Johansson, T. Duty, F. Nori and P. Delsing, Nature {\bf 479}, 376 (2011).

\bibitem{Ciuti} C. Ciuti, and I. Carusotto, Phys. Rev. A {\bf 74}, 033811 (2006).

\bibitem{Gardiner} C. W. Gardiner and M. J. Collett, Phys. Rev. A {\bf 31}, 3761 (1985).

\bibitem{GardinerBook}  C. W. Gardiner and P. Zoller, \textit{Quantum Noise}, Springer (Springer, Berlin, 2008).

\bibitem{WallsBook}  D. F. Walls and G. J. Milburn, \textit{Quantum Optics}, Springer Series in Synergetics (Springer, Berlin, 2004).

\bibitem{Delteil} A. Delteil, A. Vasanelli, Y. Todorov, C. Feuillet Palma, M. Renaudat St-Jean, G. Beaudoin, I. Sagnes, and C. Sirtori, Phys. Rev. Lett. {\bf 109}, 246808  (2012).

\bibitem{Todorov} Y. Todorov, L. Tosetto, A. Delteil, A. Vasanelli, C. Sirtori, A. M. Andrews, and G. Strasser, Phys. Rev. B {\bf 86}, 125314 (2012).

\bibitem{CohenBook}  C. Cohen-Tannoudji, J. Dupont-Roc, and G. Grynberg, \textit{Photons and Atoms: Introduction to
Quantum Electrodynamics}, (Wiley, New York, 1997).

\bibitem{Todorov3} Y. Todorov and C. Sirtori, Phys. Rev. B {\bf 85}, 045304 (2012).

\bibitem{Pegolotti} G. Pegolotti, A. Vasanelli, Y. Todorov, and C. Sirtori, Phys. Rev. B {\bf 90}, 035305 (2014).

\bibitem{Porer} M. Porer, J.-M. M\'enard, A. Leitenstorfer, R. Huber, R. Degl’Innocenti, S. Zanotto, G. Biasiol, L. Sorba, and A. Tredicucci, Phys. Rev. B {\bf 85}, 081302 (2012).

\bibitem{Riek} C. Riek, D. V. Seletskiy, A. S. Moskalenko, J. F. Schmidt, P. Krauspe, S. Eckart, S. Eggert, G. Burkard, and A. Leitenstorfer, Science {\bf 350}, 420 (2015).

\bibitem{Kawase} K. Kawase, J.-i. Shikata, H. Minamide, K. Imai, and H. Ito, Appl. Opt. {\bf 40}, 1423 (2001).

\bibitem{Edwards} T. J. Edwards, D. Walsh, M. B. Spurr, C. F. Rae, and M. H. Dunn, Opt. Express {\bf 14}, 1582 (2006).

\bibitem{Molter} D. Molter, M. Theuer, and R. Beigang, Opt. Express {\bf 17}, 6623 (2009).

\bibitem{Benea-Chelmus} I.-C. Benea-Chelmus, G. Scalari, M. Beck, and J. Faist, Phys. Rev. A {\bf 93}, 043812 (2016).

\bibitem{Bialczak} R. C. Bialczak, M. Ansmann, M. Hofheinz, M. Lenander, E. Lucero, M. Neeley, A. D. O’Connell, D. Sank, H. Wang, M. Weides, J. Wenner, T. Yamamoto, A. N. Cleland, and J. M. Martinis, Phys. Rev. Lett. {\bf 106}, 060501 (2011).

\bibitem{daSilva} M. P. da Silva, D. Bozyigit, A. Wallraff, and A. Blais, Phys. Rev. A {\bf 82}, 82, 043804 (2010).

\bibitem{Petersson} K. D. Petersson, L. W. McFaul, M. D. Schroer, M. Jung, J. M. Taylor, A. A. Houck, and J. R. Petta, Nature {\bf 490}, 380 (2012).

\bibitem{Yin} Y. Yin, Y. Chen, D. Sank, P. J. J. O’Malley, T. C. White, R. Barends, J. Kelly, E. Lucero, M. Mariantoni, A. Megrant, C. Neill, A. Vainsencher, J. Wenner,1 A. N. Korotkov, A. N. Cleland, and
J. M. Martinis, Phys. Rev. Lett. {\bf 110}, 107001 (2013).

\bibitem{Ibarcq} P. Campagne-Ibarcq, P. Six, L. Bretheau, A. Sarlette, M. Mirrahimi, P. Rouchon, and B. Huard, Phys. Rev. X {\bf 6}, 011002 (2016).

\bibitem{Yasui} T. Yasui, Y. Kabetani, E. Saneyoshi, S. Yokoyama, and T. Araki, Appl. Phys. Lett. {\bf 88}, 241104 (2006).

\bibitem{Rakic} A. D. Rakić, T. Taimre, K. Bertling, Y. L. Lim, P. Dean, D. Indjin, Z. Ikonić, P. Harrison, A. Valavanis, S. P. Khanna, M. Lachab, S. J. Wilson, E. H. Linfield and A. G. Davies, Opt. Express {\bf 21}, 22194 (2013).

\bibitem{Hoshina} H. Hoshina, T. Seta, T. Iwamoto, I. Hosako, C. Otani and Y. Kasai, J. Quant. Spectrosc. Radiat. Transf. {\bf 109}, 2303 (2008).
\end{thebibliography}
\end{document}